\documentclass[conference]{IEEEtran}
\IEEEoverridecommandlockouts
\usepackage{cite}
\usepackage{amsmath,amssymb,amsfonts}
\usepackage{algorithmic}
\usepackage{graphicx}
\usepackage{textcomp}
\usepackage{xcolor}
\usepackage{url}
\usepackage{booktabs}
\usepackage{hyperref}
\usepackage{array}
\usepackage{balance}
\def\BibTeX{{\rm B\kern-.05em{\sc i\kern-.025em b}\kern-.08em
T\kern-.1667em\lower.7ex\hbox{E}\kern-.125emX}}
\begin{document}
\raggedbottom
\bstctlcite{IEEEtranBSTCTL}

\title{AutoQResearch: LLM-Guided Closed-Loop Policy Search for Adaptive Variational Quantum Optimization}

\author{%
  \IEEEauthorblockN{Monit SHARMA}
  \IEEEauthorblockA{\textit{School of Computing and Information Systems} \\
    \textit{Singapore Management University}\\
    Singapore \\
  monitsharma@smu.edu.sg}
  \and
  \IEEEauthorblockN{Hoong Chuin LAU\textsuperscript{\dag}\thanks{\dag Corresponding Author}}
  \IEEEauthorblockA{\textit{School of Computing and Information Systems} \\
    \textit{Singapore Management University}\\
    Singapore \\
  hclau@smu.edu.sg}
}

\maketitle

\begin{abstract}
  Configuring variational quantum algorithms for combinatorial optimization remains
  a difficult and largely expert-driven process, requiring coordinated choices over
  solver family, ansatz, objective, optimizer, encoding, and fallback logic. We
  present \textit{AutoQResearch}, an LLM-guided closed-loop experimentation framework
  that casts this task as sequential policy search over a curated quantum-solver
  design space. Instead of searching for a single static configuration, the framework
  searches for adaptive solver-control policies that condition future decisions on
  diagnostics from earlier attempts, including feasibility, optimality gap,
  convergence stagnation, and sampling concentration.

  The system operates through a structured and auditable workflow: an LLM agent edits
  a small policy surface under a fixed evaluation harness, candidate policies are first
  screened using cheap scout evaluations, and only the strongest candidates are promoted
  to full confirmation. The code and experiment artifacts are publicly released to
  support reproduction of the search traces, plots, and policy checkpoints.\footnote{Code
  and artifacts: \url{https://github.com/SMU-Quantum/autoqresearch}} This enables
  controlled autonomous exploration while guarding against proxy overfitting and unstable
  policy selection.

  We evaluate the framework on two combinatorial optimization settings with different
  structural characteristics: Maximum Independent Set (MIS) and the Capacitated Vehicle
  Routing Problem (CVRP). On MIS instances from 16 to 64 vertices, the discovered
  policies substantially outperform the static baseline and reveal clear
  scale-dependent solver behavior: CVaR objectives are most effective at small scale,
  warm-start QAOA transfers well at 16 nodes but collapses at 32, and QRAO-based qubit
  compression provides the most effective explored scaling path. On CVRP curricula from
  8- to 12-customer instances and a held-out E-n13-k4 benchmark, the framework
  discovers policy adaptations involving sampling budget, penalty design, and hybrid
  feasibility-repair protocols, yielding feasible high-quality solutions on held-out
  instances.

  Beyond the problem-specific results, the main methodological finding is that staged
  confirmation is essential: cheap proxy evaluations can materially misestimate policy
  quality and even invert candidate rankings. Overall, the paper positions AutoQResearch
  as a benchmarked quantum--GenAI co-design workflow for autonomous solver discovery in
  variational quantum optimization.
\end{abstract}


\section{\label{sec:level1}Introduction}

Recent interest in quantum--generative AI co-design has highlighted the need for
closed-loop workflows in which generative models do more than summarize results or
assist with coding: they must propose candidates, interact with computational
backends, interpret diagnostic feedback, and refine future actions under an auditable
evaluation protocol. Such workflows are especially relevant in variational quantum
optimization, where performance depends on a large number of interacting design
choices and where useful feedback is naturally produced after each solver attempt.

Variational quantum algorithms (VQAs), including the Variational Quantum Eigensolver
(VQE) \cite{Peruzzo_2014}, the Quantum Approximate Optimization Algorithm (QAOA)
\cite{farhi2014quantumapproximateoptimizationalgorithm}, and related variants such as
CVaR-VQE \cite{Barkoutsos_2020}, Warm-Start QAOA \cite{Egger_2021},
Multi-Angle QAOA \cite{herrman2021multianglequantumapproximateoptimization},
Quantum Random Access Optimization (QRAO)
\cite{fuller2021approximatesolutionscombinatorialproblems,10821401}, and Pauli Correlation
Encoding (PCE) \cite{Sciorilli_2025, sharma2025comparativestudyquantumoptimization}, expose a rich and highly coupled solver design
space. Practical performance depends jointly on solver family, ansatz architecture,
depth, classical optimizer, measurement objective, encoding strategy, compression
ratio, and fallback behavior. These choices interact with problem structure, instance
scale, and resource constraints in ways that are difficult to predict analytically,
making solver configuration a major practical bottleneck.

Most existing tuning approaches, such as grid search, random search, and Bayesian
optimization \cite{Shah}, treat solver design as a static hyperparameter-selection
problem. In contrast, quantum optimization is inherently sequential and
diagnostic-rich: each attempt produces information about feasibility, objective
quality, convergence behavior, and sample concentration that can inform the next
decision. This suggests that the appropriate object of optimization is not merely a
single fixed configuration, but an adaptive decision policy that maps observed solver
diagnostics to subsequent actions.

In this paper, we introduce \textit{AutoQResearch}, an LLM-guided closed-loop
experimentation framework that casts variational quantum algorithm configuration as
sequential policy search over a curated and auditable solver design space. Rather than
allowing unrestricted code generation, the framework restricts the agent to editing a
small policy surface under a fixed evaluation harness. Candidate policies are
proposed, executed, screened using cheap scout evaluations, and then promoted to
full confirmation only if they remain competitive under broader testing. In this
sense, the system functions as a quantum--GenAI co-design workflow: the generative
model proposes adaptive solver-control logic, and the quantum optimization stack
validates, refines, or rejects it through measured execution outcomes.

We evaluate the framework on two combinatorial optimization settings with different
structural characteristics. The first is Maximum Independent Set (MIS), which provides
a clean constrained QUBO benchmark for studying solver-family switching, objective
selection, and compression-based scaling. The second is a decomposed Capacitated
Vehicle Routing Problem (CVRP) workflow, which introduces a richer decision pipeline
involving assignment feasibility, penalty design, sampling-budget allocation, and
repair-aware evaluation. This dual evaluation allows us to study not only whether the
agent can discover effective policies, but also whether the same closed-loop search
framework remains useful across qualitatively different quantum optimization regimes.

\subsection{Why Adaptive Policies Instead of Static Configurations?}

A static solver configuration cannot express conditional logic of the form:
``if the current attempt is feasible but stagnant, change the measurement objective,''
or ``if the current family loses concentration at larger scale, switch to a compressed
encoding and use a fallback rule.'' Yet these are precisely the kinds of responses
that arise in practice during variational quantum optimization. In our experiments,
different failure modes required different interventions, including changes in
measurement strategy, solver family, compression ratio, penalty design, and sampling
budget. This motivates viewing the design problem as sequential policy search rather
than static hyperparameter tuning.

\subsection{LLM Guidance Within a Structured Search Workflow}

Recent work on LLM-assisted scientific discovery suggests that language models can
support iterative experimental workflows when they are embedded within structured
execution environments \cite{Boiko2023,lu2024aiscientistfullyautomated,huang2024mlagentbenchevaluatinglanguageagents,karpathy2026autoresearch}. Our setting adopts this perspective, but under a deliberately restricted design. The LLM is not asked to invent new quantum algorithms from scratch. Instead, it acts as a sequential experiment planner over a curated solver space, editing a small policy surface while operating under a fixed evaluation harness and explicit promotion rules. This restriction is intentional: it makes the search process auditable, reproducible, and scientifically interpretable.

A natural question is whether the proposed framework offers advantages over classical search strategies such as Bayesian optimization, or bandit-based controllers over the same policy space. These methods are effective when the search space is low-dimensional, fixed, and smoothly parameterized.
In contrast, the policy space in AutoQResearch is \emph{structured, discrete, and programmatic}: it includes solver-family switching, objective selection, conditional fallback logic, and workflow-level transformations that depend on interpreting heterogeneous diagnostic signals. This space is not naturally
amenable to continuous optimization or simple surrogate modeling. The role of the LLM in this setting is not merely to sample configurations, but to perform
\emph{diagnosis-driven search}, proposing targeted, semantically meaningful policy edits conditioned on observed failure modes (e.g., stagnation, infeasibility, or concentration collapse). This enables rapid navigation of a combinatorial design space that would otherwise require extensive manual engineering or prohibitively large numbers of blind evaluations.

Conceptually, AutoQResearch sits at the intersection of three classical lineages:
automated algorithm configuration and AutoML, in which solver settings are tuned by
random search or Bayesian optimization~\cite{Shah}; per-instance algorithm selection and
solver portfolios, which map instance features to a chosen solver; and agentic
experimentation systems, in which an LLM plans and executes iterative experiments under
tool feedback~\cite{Boiko2023,lu2024aiscientistfullyautomated,huang2024mlagentbenchevaluatinglanguageagents,karpathy2026autoresearch}.
Our contribution relative to the first two is that the search object is an
\emph{adaptive control policy} over a discrete, programmatic space---including
solver-family switches and conditional fallback logic---rather than a point in a fixed
continuous configuration space; relative to the third, the agent is deliberately confined
to a small typed policy surface behind a fixed harness, trading open-ended autonomy for
auditability and reproducibility.

\subsection{Contributions}

We make the following contributions:

\begin{itemize}

  \item \textbf{A closed-loop quantum--GenAI framework for adaptive solver discovery.}
    We introduce \textit{AutoQResearch}, an LLM-guided experimentation framework that
    casts variational quantum solver design as sequential policy search over a curated
    and auditable action space, rather than as static hyperparameter selection.

  \item \textbf{A structured and reproducible agent interface for quantum optimization.}
    The framework restricts the agent to a small editable policy surface under a fixed
    evaluation harness, enabling autonomous yet controlled exploration of solver-family
    choice, objective selection, circuit design, compression strategy, sampling budget,
    and fallback logic.

  \item \textbf{A staged evaluation methodology for reliable autonomous search.}
    We propose a scout--promote--confirm protocol with replay guardrails, showing that
    cheap proxy evaluations can materially misestimate policy quality and even invert
    candidate rankings. This provides a practical evaluation methodology for closed-loop
    quantum--GenAI workflows.

  \item \textbf{Cross-problem evidence on adaptive policy discovery.}
    We evaluate the framework on two structurally different optimization settings:
    Maximum Independent Set (MIS) and a decomposed Capacitated Vehicle Routing Problem
    (CVRP) workflow. Across these settings, the agent discovers nontrivial problem- and
    scale-dependent policy adaptations, including objective switching, solver-family
    changes, compression-based scaling, penalty redesign, and sampling-budget control.

  \item \textbf{Open and auditable research artifacts.}
    We release the framework, benchmark instances, policy checkpoints, experiment logs,
    and analysis scripts to support reproducibility and future work on autonomous
    quantum optimization workflows.

\end{itemize}

\section{Problem Formulation}

AutoQResearch is designed to search over adaptive solver-control policies for
variational quantum optimization workflows. In this paper, we evaluate the framework
on two combinatorial optimization settings with different structural properties:
Maximum Independent Set (MIS) and a decomposed Capacitated Vehicle Routing Problem
(CVRP) workflow. The purpose of using both is not to introduce new mathematical
formulations for these problems, but to test whether the same closed-loop
quantum--GenAI search procedure can discover effective policies across distinct
optimization regimes.

\subsection{Maximum Independent Set}

The Maximum Independent Set (MIS)~\cite{mis} problem on an undirected graph
$G=(V,E)$ seeks the largest subset of vertices such that no two selected vertices
are adjacent:
\[
  \begin{aligned}
    \max_{x \in \{0,1\}^{|V|}} \quad & \sum_{i \in V} x_i \\
    \text{subject to} \quad & x_i + x_j \le 1, \quad \forall (i,j)\in E.
  \end{aligned}
\]

MIS is NP-hard and is a standard benchmark for variational quantum optimization
because feasibility itself becomes nontrivial once adjacency constraints are encoded
through soft penalties. In our framework, the constrained problem is converted into
a QUBO of the form
\[
  \min_x \left(
    -\sum_i x_i + P \sum_{(i,j)\in E} x_i x_j
  \right),
\]
where $P>0$ is a penalty coefficient. Since each selected vertex contributes reward
$1$ while each violated edge must be discouraged, feasibility requires $P>1$; the
penalty is auto-derived by the QUBO conversion so that any single constraint violation
strictly outweighs the maximum achievable objective reward, which for unit-weight MIS
yields $P\ge 2$. This coefficient is exposed to the agent as a tunable degree of
freedom (via the \texttt{penalty} field), but on the reported runs the agent left it at
the auto-derived value. The corresponding Ising Hamiltonian is then solved using one of
the supported quantum solver families.

\subsection{Decomposed CVRP Workflow}

The Capacitated Vehicle Routing Problem (CVRP)~\cite{toth2002vehicle} requires a fleet of capacitated
vehicles to serve customer demands from a depot at minimum travel cost. Rather than
treating CVRP as a single monolithic quantum optimization problem, our experiments
use a decomposed workflow~\cite{sharma2026qubitscalablecvrplagrangianknapsack,cvrp_monit} in which a quantum solver is applied to an intermediate
assignment subproblem and a classical routing procedure evaluates the resulting
clusters.

More precisely, each CVRP instance is first transformed into a seed-based
customer-to-vehicle assignment problem inspired by Fisher--Jaikumar-style
decomposition~\cite{fisher1981generalized}. This assignment stage determines how customers are partitioned across
vehicles subject to capacity considerations. The resulting assignment problem is then
encoded as a QUBO and solved by the candidate quantum policy. Given a decoded
assignment, route costs are computed by a fixed downstream routing stage, allowing
the overall pipeline to be scored by final CVRP solution quality.

This decomposed setting is useful for AutoQResearch because it exposes a different
policy-search regime from MIS\@. In addition to solver-family and objective choices,
performance depends on assignment feasibility, penalty design, seed selection,
sampling budget, and repair behavior. As a result, CVRP provides a complementary
test of whether the agent can adapt its decisions in a richer hybrid
quantum-classical workflow.

A key adaptation discovered on this workflow concerns how the vehicle-capacity
constraint is encoded. Writing $s_v = \sum_{c} d_c\, y_{cv} - Q_v$ for the load surplus
of vehicle $v$ (with customer demands $d_c$, assignment variables $y_{cv}$, and capacity
$Q_v$), the initial encoding uses a symmetric quadratic slack penalty $\lambda\, s_v^2$
that charges both under- and over-capacity deviations. The \emph{tilted} penalty replaces
this with an asymmetric, one-sided form that charges only capacity \emph{overflow},
$\lambda\,[\max(0,s_v)]^2$, optionally with a stronger multiplier on the overflow branch.
This removes the spurious pressure that the symmetric form places on legitimately
under-full vehicles, enlarging the feasible region sampled by the quantum solver. The
penalty weight $\lambda$ is scaled relative to the assignment reward; exact coefficients
for each stage are provided in the released configuration.

\subsection{Evaluation Objective}

For both problem settings, the search target is solution quality measured relative to
a classical reference. On MIS, we use per-instance optimality gap relative to the
exact maximum independent set size, assigning gap $1.0$ to infeasible outputs. On
CVRP, we use route-cost gap relative to the benchmark reference solution generated by
the fixed evaluation pipeline, again treating infeasible decoded assignments as worst
case. Candidate policies are compared through suite-level average gap over the
relevant curriculum stage, and this quantity is the optimization target throughout
the search.

To avoid ambiguity, we state explicitly where feasibility is evaluated, since it plays
several roles. (i) As an \emph{agent-facing diagnostic}, per-attempt feasibility and
feasibility rate are reported to the controller so it can react to infeasibility, but
they never directly enter the objective. (ii) In the \emph{MIS gap}, the best sampled
bitstring is checked against the adjacency constraints; an infeasible best sample yields
gap $1.0$, so feasibility is folded into the single reported quantity. (iii) In the
\emph{CVRP pipeline}, feasibility is checked on the decoded assignment; a bounded
classical repair step may restore capacity feasibility before routing, and only the
repaired feasible assignment is routed and scored (an assignment that cannot be repaired
is scored as worst case). In all cases the number that drives keep/revert and that we
report is the feasibility-aware suite-average gap; feasibility is a gate on that gap, not
a separate reward term.

\section{Methodology}

\subsection{Closed-Loop Search Workflow}

AutoQResearch treats variational quantum solver design as a closed-loop sequential
search problem. At each iteration, an LLM agent proposes edits to a restricted policy
surface that controls how a solver is selected, configured, and adapted across
attempts. The edited policy is then executed by a fixed evaluation harness on the
current benchmark stage, which returns quantitative diagnostics including feasibility,
solution quality, and sampling behavior. These outcomes are used both to score the
candidate policy and to inform the next round of policy edits.

This design deliberately separates \emph{proposal} from \emph{validation}. The LLM
does not directly decide which candidate is accepted; rather, it proposes a policy,
the quantum optimization stack evaluates it under fixed rules, and the measured
results determine whether the candidate is retained, discarded, or promoted for
broader confirmation. In this sense, the framework implements a closed-loop
quantum--GenAI workflow in which the generative model proposes adaptive
solver-control logic and the solver backend validates, refines, or rejects it.

\subsection{Policy Interface and Observation Signals}

The agent operates over a small editable policy surface implemented through four
functions: selecting the initial solver family, constructing a base configuration,
deciding whether another attempt should be executed, and adapting the policy after
observing previous outcomes. This interface is expressive enough to represent both
static configurations and conditional controllers with fallback logic, while
remaining compact and auditable.

After each attempt, the controller receives a structured outcome summary containing
the optimization target (gap), feasibility information, and a small set of
diagnostic signals used throughout the search. In the MIS setting, the most relevant
signals include raw feasibility, feasibility rate, convergence stagnation, and the
probability mass of the most likely sampled bitstring. In the CVRP workflow, the
same interface is used, but the dominant signals additionally reflect assignment
feasibility and downstream routed solution quality. These observations allow the
agent to condition future actions on failure modes such as stagnation, infeasibility,
or concentration collapse.

\subsection{Agent Configuration and Policy Space}
\label{sec:agent_config}

\paragraph{LLM and prompting.}
We use \textbf{GPT-5.3-Codex} with identical instructions for both the MIS and
CVRP studies. The complete natural-language prompt is the released
instruction file \texttt{program.md}; it fixes the objective (minimize suite-average
gap, lower is better), the four editable policy functions, the available degrees of
freedom, and a mandatory protocol (at least fifteen search iterations, the seed treated
as a verification-only control, and classical local search disabled). At each iteration
the model receives the current policy source together with the structured diagnostic
summary of the previous attempt, and it may edit only \texttt{experiment.py} (the four
policy functions) and its research log \texttt{agent\_journal.md}. No fine-tuning is
performed and no gradient signal is passed to the model: the only feedback is the textual
diagnostic summary and the keep/revert decision returned by the harness. Cross-stage
memory is provided by curriculum carry-forward---the best confirmed snapshot of a stage
is restored before the next stage begins---and by the agent's own journal, rather than by
any hidden model state. The prompt and protocol are held constant across the two problem
settings; only the benchmark suites differ.

\paragraph{Policy space and how it was designed.}
The editable action space is discrete, programmatic, and deliberately human-curated so
that every proposal is auditable. It contains four solver families (VQE, QAOA, PCE,
QRAO); per-family choices of ansatz (\texttt{real\_amplitudes}, \texttt{efficient\_su2},
\texttt{pauli\_two\_design}, brickwork, or a custom builder), circuit depth/repetitions,
entanglement pattern, classical optimizer and its budget, measurement objective
(expectation or CVaR with tunable $\alpha$), estimator/sampler shot counts, and the QUBO
penalty; the QAOA variant (standard, warm-start with mixing $\varepsilon$, or
multi-angle); and the QRAO compression ratio ($1{:}1$, $2{:}1$, $3{:}1$) with
semideterministic or magic rounding. Crucially, the surface also exposes the
\emph{control flow}: a stopping rule and an adaptation rule that may change any of the
above \emph{between attempts on a single instance}, conditioned on observed diagnostics.
The space was designed by the standard degrees of freedom of the included solver
families; the guiding principle for admitting an action was that it should let the agent
respond to a diagnosable failure mode (poor concentration, low feasibility, optimizer
stagnation, runtime blow-up, or size-specific failure). Two exclusions are enforced by
design: classical local search is disabled (\texttt{pce\_local\_search} and
\texttt{final\_local\_search} fixed to \texttt{False}) so that families compete purely on
quantum output, and the random seed is fixed ($=17$) and explicitly forbidden as a tuning
knob so that improvements reflect algorithmic change rather than a favorable draw. Even
restricting to discrete choices and ignoring the conditional control flow, a single
attempt admits on the order of $10^{3}$--$10^{4}$ configurations; because an adaptive
policy is a branching program over such attempts, the effective search space is not
enumerable by grid or random search at the budgets used here, which is the sense in which
the agent performs diagnosis-driven search rather than enumeration.

\paragraph{Reliability and guardrails.}
Because the agent edits only a small typed policy surface behind a fixed harness, a
proposal cannot alter the evaluation or the metric. On the MIS trajectory analyzed here
the agent issued $17$ policy proposals (in addition to the recorded baseline); all $17$
executed to completion with no runtime error, $9$ were kept and $8$ were reverted, and
every keep/revert decision was made by the harness on the measured suite metric under a
strict-improvement rule with a $0.02$ dev/replay regression tolerance. Reverts restore
the previous git snapshot, so a dominated or self-contradictory edit cannot persist. In
this restricted setting we observed no executable-but-invalid ``hallucinated'' policies;
the operative failure mode is not malformed code but proposals that fail to improve the
metric, which the guardrails reject automatically. We regard this containment as a
feature of the restricted-surface design rather than evidence that the underlying model
never errs.

\subsection{Staged Evaluation Protocol}

Evaluating every candidate policy on the full benchmark at each iteration would be
unnecessarily expensive and would make the search unstable. We therefore use a
three-stage evaluation protocol.

\paragraph{Scout.}
A candidate policy is first evaluated on a cheap proxy consisting of a small subset
of instances from the current stage, together with replay checks on earlier stages
when required. This provides rapid feedback for search.

\paragraph{Promote.}
Only the strongest scout candidates are re-evaluated on the full stage benchmark.
This filters out candidates that overfit the proxy or benefit from favorable noise.

\paragraph{Confirm and carry forward.}
The best promoted candidate becomes the locked controller for the current stage and
is then carried forward to the next stage in the curriculum. Replay guardrails are
applied so that improvements at a new stage do not come at the cost of large
regressions on earlier validated stages.

This protocol is a central methodological component of the framework. In our
experiments, cheap proxy evaluations were sometimes overly optimistic and could even
invert candidate rankings, making explicit confirmation necessary for reliable
autonomous search.

\subsection{Benchmarks and Solver Families}

We evaluate the framework on curriculum-based benchmark ladders for MIS and CVRP\@.
For MIS~\cite{sharma2025comparativestudyquantumoptimization,mis}, the curriculum spans graph instances from 16 to 64 vertices. For CVRP~\cite{sharma2026qubitscalablecvrplagrangianknapsack,cvrplib}, the
curriculum progresses from 8-customer to 12-customer decomposed assignment-routing
instances, followed by a held-out E-n13-k4 benchmark. At each stage, the search
objective is the suite-average gap over the active benchmark split.

The available solver families are drawn from a curated quantum optimization design
space including VQE, QAOA, QRAO, and PCE, together with solver-specific choices such
as ansatz type, circuit depth, optimizer, measurement objective, compression ratio,
rounding method, and sampling budget. In the CVRP workflow, the search space also
includes decomposition-sensitive controls such as seed selection, penalty design, and
repair-aware sampling decisions. All candidates are executed under a fixed harness so
that policy comparisons remain controlled and reproducible.

\paragraph{Instances and splits.}
Each curriculum stage has an explicit split. Search is driven by a cheap \emph{scout}
proxy (two instances per stage); a stage winner is declared only on the fuller
\emph{confirm} suite (five instances at 16 and 32 nodes); and earlier stages are
re-evaluated as \emph{replay} guardrails whose gap may not worsen by more than $0.02$.
The 48-node stage uses a retained sparse instance, and the 64-node instance is
\emph{held out}: it is never used for keep/discard and is scored only once the final
policy is locked. MIS instances are drawn from the benchmark set
of~\cite{sharma2025comparativestudyquantumoptimization,mis}, and the held-out CVRP
instance is \texttt{E-n13-k4} from CVRPLIB~\cite{cvrplib}. Exact per-stage instance
membership, counts, and sources are released with the artifact.

\paragraph{Simulation backend.}
All policy search and confirmation runs use \emph{ideal, noiseless} simulation: a
matrix-product-state / statevector backend (harness mode \texttt{ideal\_mps}, with
\texttt{statevector} as an alias) with fixed estimator and sampler shot budgets.
Noisy-simulator and hardware backends are supported by the same harness but are reserved
for the hardware validation of Section~\ref{sec:hardware}. Policies are therefore
\emph{discovered under idealized simulation}, isolating algorithmic signal from device
noise; the hardware runs then test whether the discovered policies survive transfer to
real devices rather than being tuned to a particular noise model. Unless stated
otherwise the seed is fixed at $17$; a post-lock robustness protocol re-evaluates the
final held-out policy over five seeds ($17,23,29,31,37$).

\section{Results}
\label{sec:results}

We report results on two problem settings: Maximum Independent Set (MIS) and a
decomposed Capacitated Vehicle Routing Problem (CVRP) workflow. In both settings,
the objective of the search is to minimize suite-average gap under a staged
scout--promote--confirm protocol. The results are presented in a compact form to
highlight two complementary aspects of the framework: \emph{search dynamics}, shown
through stage-wise trajectory plots, and \emph{validated outcomes}, shown through
curriculum-level summary tables and overview figures.

\subsection{Case Study I: Maximum Independent Set}

Table~\ref{tab:mis_stage_summary} summarizes the final MIS controllers discovered
across the curriculum. At 16 nodes, the strongest confirmed controller is
warm-start QAOA with CVaR, which improves the confirmed stage gap to $0.250$ from a
substantially weaker static baseline. At 32 nodes, the search shifts away from
direct QAOA transfer and identifies QRAO with 3:1 compression and magic rounding as
the strongest confirmed controller, achieving a gap of $0.247$. In the larger-size
regime, the locked controller becomes adaptive: a semideterministic QRAO policy with
fallback from 3:1 to 2:1 compression yields a confirmed gap of $0.533$ on the
retained 48-node instance and transfers to the unseen 64-node held-out instance with
gap $0.550$. These results show that the discovered controller is not uniform across
scale; instead, the effective policy changes from CVaR-enhanced QAOA at small scale
to qubit-compressed QRAO variants at larger scale.

\begin{figure*}[t]
  \centering
  \includegraphics[width=\linewidth]{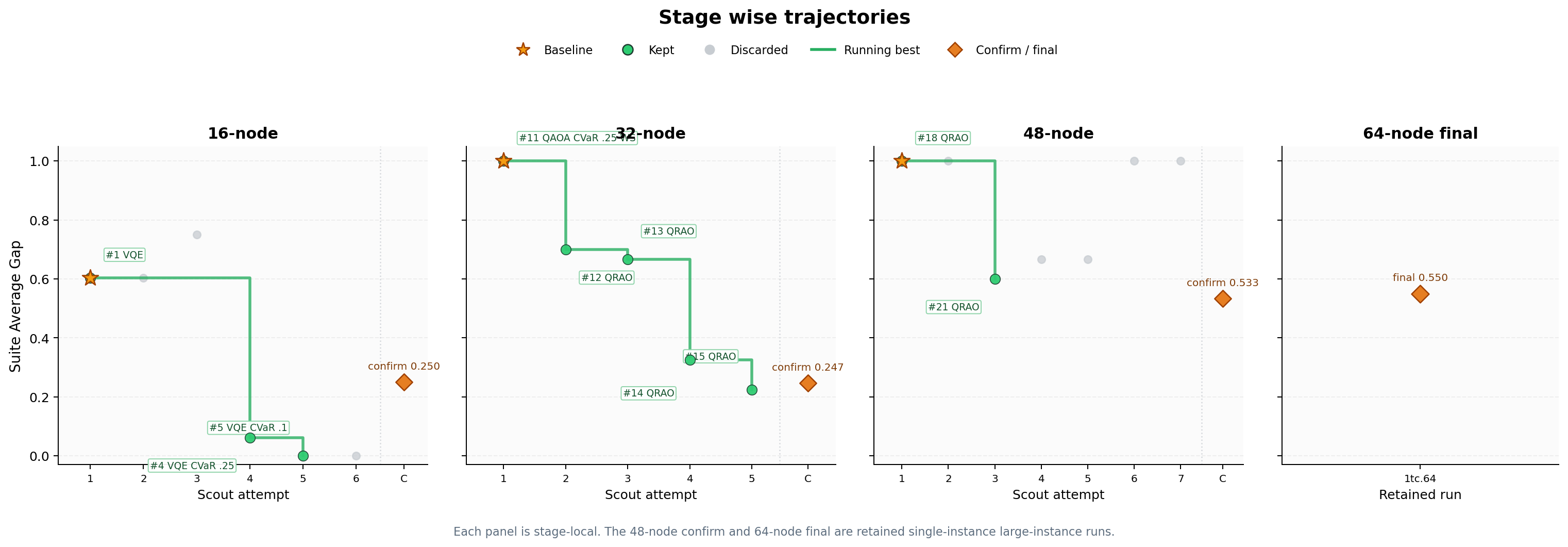}
  \caption{Stage-wise MIS search trajectories. Each panel is local to one stage.
    Green points denote retained scout candidates, gray points denote discarded
    candidates, stars denote stage baselines, and orange diamonds denote confirmed
    winners or the held-out final. The trajectories show that the best solver family
    changes with scale: CVaR-enhanced warm-start QAOA is strongest at 16 nodes,
  whereas QRAO-based compressed policies dominate at larger sizes.}
  \label{fig:mis_progress}
\end{figure*}

Figure~\ref{fig:mis_progress} shows the stage-local search trajectories. The 16-node
stage begins with a feasible but weak VQE baseline, after which the search discovers
that changing the objective is more valuable than increasing circuit complexity.
Warm-start QAOA with CVaR then becomes the confirmed winner. At 32 nodes, the same
family no longer transfers, and the search moves toward QRAO-based compressed
controllers. At 48 nodes, no static single-shot strategy remains reliable, and the
final controller uses conditional fallback logic. This figure therefore illustrates
the central mechanism of AutoQResearch: the agent does not simply optimize a single
configuration, but adapts its strategy as the dominant failure mode changes with
scale.

\begin{figure}[t]
  \centering
  \includegraphics[width=\linewidth]{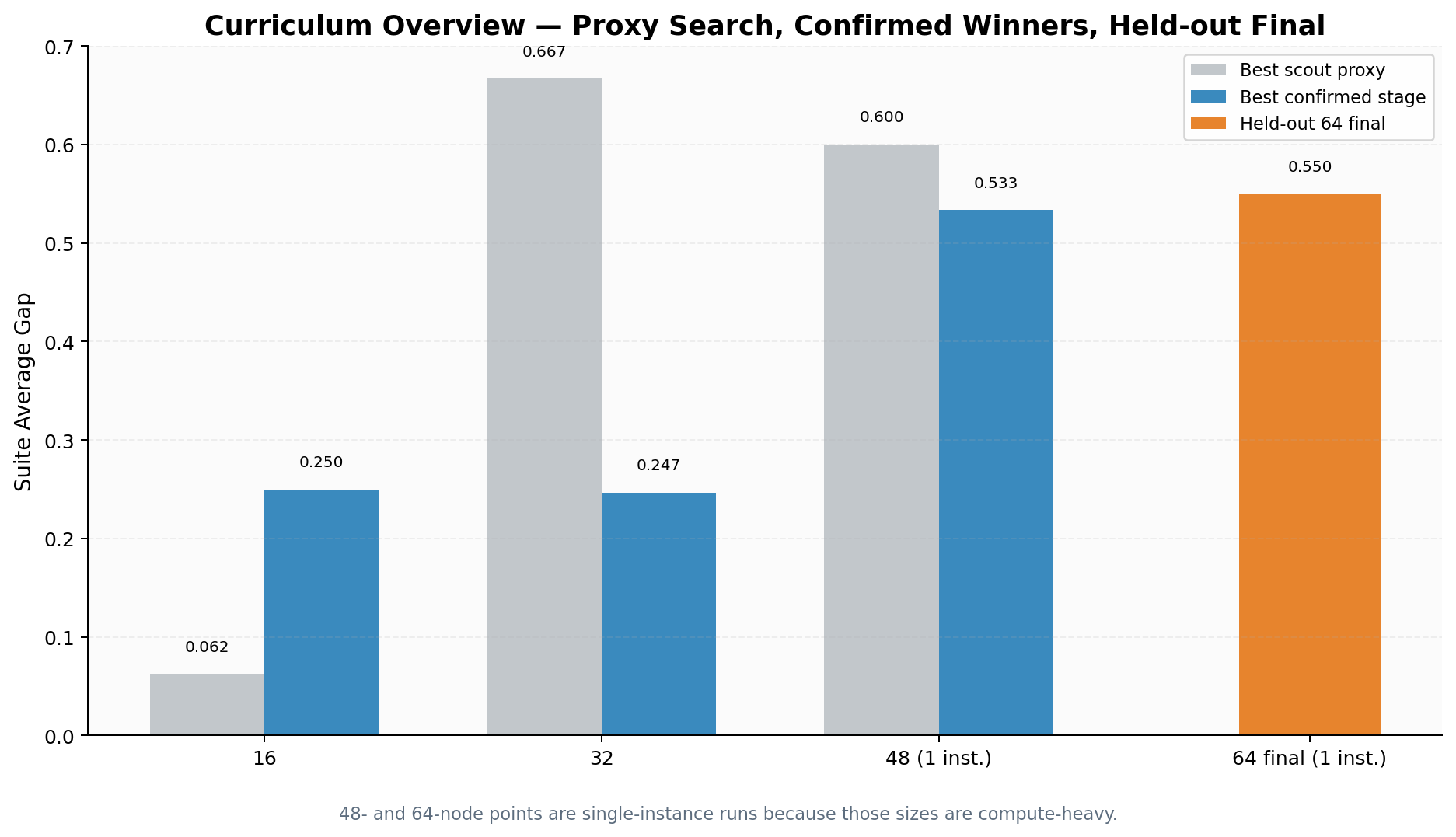}
  \caption{MIS curriculum overview comparing best scout proxy, best confirmed stage,
    and the held-out 64-node final. The discrepancy between scout and confirmed
    results shows that proxy-only selection is unreliable, motivating the staged
  scout--promote--confirm protocol.}
  \label{fig:mis_overview}
\end{figure}

Figure~\ref{fig:mis_overview} summarizes the relationship between proxy and confirmed
performance. Two patterns are clear. First, scout results can be materially
optimistic relative to confirmed results. Second, the identity of the strongest proxy
candidate need not match the strongest confirmed policy, especially at intermediate
scales. This validates the need for the promotion step in the closed-loop search
workflow and supports our claim that reliable autonomous experimentation requires
staged confirmation rather than proxy-only search.

\begin{table*}[t]
  \centering
  \caption{Summary of discovered MIS controllers across curriculum stages.}
  \label{tab:mis_stage_summary}
  \begin{tabular}{lccc}
    \toprule
    \textbf{Stage} & \textbf{Best policy} & \textbf{Confirmed/final gap} & \textbf{Role} \\
    \midrule
    16-node & WS-QAOA + CVaR($\alpha=0.25$) & 0.250 & confirmed \\
    32-node & QRAO 3:1 + magic rounding & 0.247 & confirmed \\
    48-node & Adaptive QRAO 3:1 $\rightarrow$ 2:1 & 0.533 & confirmed \\
    64-node & Adaptive QRAO 3:1 $\rightarrow$ 2:1 & 0.550 & held-out final \\
    \bottomrule
  \end{tabular}
\end{table*}

\subsection{Staged-Confirmation Ablation}
\label{sec:ablation}

The claim that staged confirmation is \emph{necessary} rather than merely convenient can
be tested directly from the promotion logs, which record both the cheap scout-proxy gap
and the full-suite confirmed gap for every promoted candidate. Table~\ref{tab:ablation}
contrasts, within one promotion round at each MIS stage, the policy a \emph{scout-only}
rule would ship (the scout-argmin) against the policy selected under confirmation (the
confirm-argmin).

At 16 nodes the two-instance scout proxy saturates: the scout-best candidate reaches gap
$0.000$ on the proxy yet confirms at $0.400$ on the full five-instance suite, while a
candidate the proxy ranked \emph{worse} (scout $0.063$) confirms at $0.250$. A scout-only
rule would therefore ship a policy $0.150$ worse than the confirmed choice. At 32 nodes
the inversion is sharper: the confirm-best candidate (confirmed gap $0.247$) was ranked
\emph{last of three} by the scout proxy (scout gap $0.667$), so any top-1 scout rule would
have \emph{discarded} the eventual winner and shipped a policy with confirmed gap $0.292$.
In both rounds the proxy ordering does not match the confirmed ordering---exactly the
failure the promote/confirm stage exists to catch. Throughout the 32-node search the
16-node replay guardrail held at gap $0.025$, confirming that the advance did not silently
regress the earlier stage. For context on cost, the full MIS search consumed $17$ policy
proposals and roughly $4.1$ hours of cumulative simulation across scout, confirm, and
replay evaluations, with zero execution failures. These figures give a concrete,
log-derived basis for the paper's central methodological claim; a matched-budget
head-to-head against non-LLM search strategies over the same surface remains future work
(Section~\ref{sec:discussion}).

\begin{table}[t]
  \centering
  \caption{Staged-confirmation ablation on MIS, from the recorded promotion rounds. Each
    row is a promoted candidate with its cheap scout-proxy gap and its full-suite confirmed
    gap (lower is better). A scout-only rule would select the scout-argmin
    (\underline{underlined}); confirmation selects the confirm-argmin (\textbf{bold}). At
    both stages the two disagree, and at 32 nodes the confirmed winner is the
  worst scout-ranked candidate---so a proxy-only rule would have discarded it.}
  \label{tab:ablation}
  \begin{tabular}{llcc}
    \toprule
    \textbf{Stage} & \textbf{Promoted candidate} & \textbf{Scout gap} & \textbf{Confirm gap} \\
    \midrule
    16-node & VQE + CVaR($\alpha{=}0.10$) & \underline{0.000} & 0.400 \\
    16-node & VQE + CVaR($\alpha{=}0.25$) & 0.063 & \textbf{0.250} \\
    16-node & VQE baseline               & 0.604 & 0.581 \\
    \midrule
    32-node & QRAO $3{:}1$ (a)            & \underline{0.225} & 0.292 \\
    32-node & QRAO $3{:}1$ (b)            & 0.325 & 0.450 \\
    32-node & QRAO $3{:}1$ (c)            & 0.667 & \textbf{0.247} \\
    \bottomrule
  \end{tabular}
\end{table}

\subsection{Case Study II: Decomposed CVRP Workflow}

Table~\ref{tab:cvrp_stage_summary} summarizes the final CVRP results. Unlike MIS,
where solver-family switching is the dominant adaptation axis, the CVRP workflow
reveals a different regime: performance depends strongly on assignment feasibility,
penalty design, and sampling budget. On the 8-customer stage, the search improves the
confirmed gap from a baseline scout gap of $0.112$ to a confirmed stage gap of
$0.095$. On the 10-customer stage, the strongest confirmed policy achieves a gap of
$0.047$, driven by the transition from hard-slack capacity handling to a tilted
penalty formulation~\cite{sharma2026qubitscalablecvrplagrangianknapsack} together with higher final sampling. On the retained 12-customer
run, the hybrid protocol yields a feasible result with gap $0.010$. Finally, on the
held-out E-n13-k4 benchmark, the locked policy attains a feasible final gap of
$0.139$. These results show that the same AutoQResearch framework remains effective
in a richer hybrid routing pipeline, but the useful interventions differ from those
observed for MIS.

\begin{figure*}[t]
  \centering
  \includegraphics[width=\linewidth]{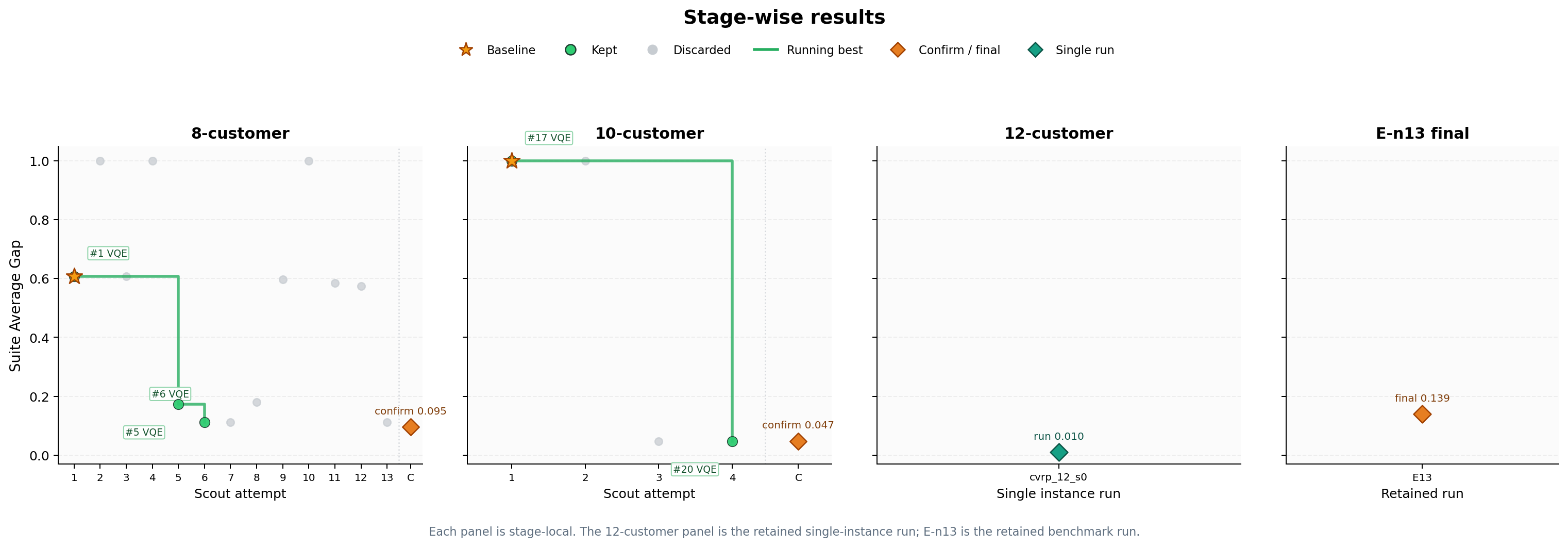}
  \caption{Stage-wise CVRP search trajectories. The 8-customer stage improves
    feasibility through increased sampling, while the 10-customer stage improves
    sharply after adopting a tilted capacity penalty. The retained 12-customer run
    and held-out E-n13 benchmark demonstrate that the framework can discover useful
  workflow-level adaptations beyond the initial curriculum stages.}
  \label{fig:cvrp_progress}
\end{figure*}

Figure~\ref{fig:cvrp_progress} shows the CVRP search trajectory. At 8 customers, the
early search establishes that pure solver-family changes are less important than
recovering rare feasible assignment samples. At 10 customers, the dominant
improvement comes from changing the underlying assignment construction, specifically
the move to tilted capacity penalties, which makes both training instances feasible
while preserving replay performance. The retained 12-customer run and the held-out
E-n13 benchmark then demonstrate that the resulting policy is not confined to the
earlier curriculum stages. In contrast to the MIS study, where compression and
family-switching dominate the larger-size regime, the CVRP study highlights the role
of workflow-level decisions such as penalty redesign, replay-aware selection, and
feasibility-repair-aware evaluation.

\begin{figure}[t]
  \centering
  \includegraphics[width=\linewidth]{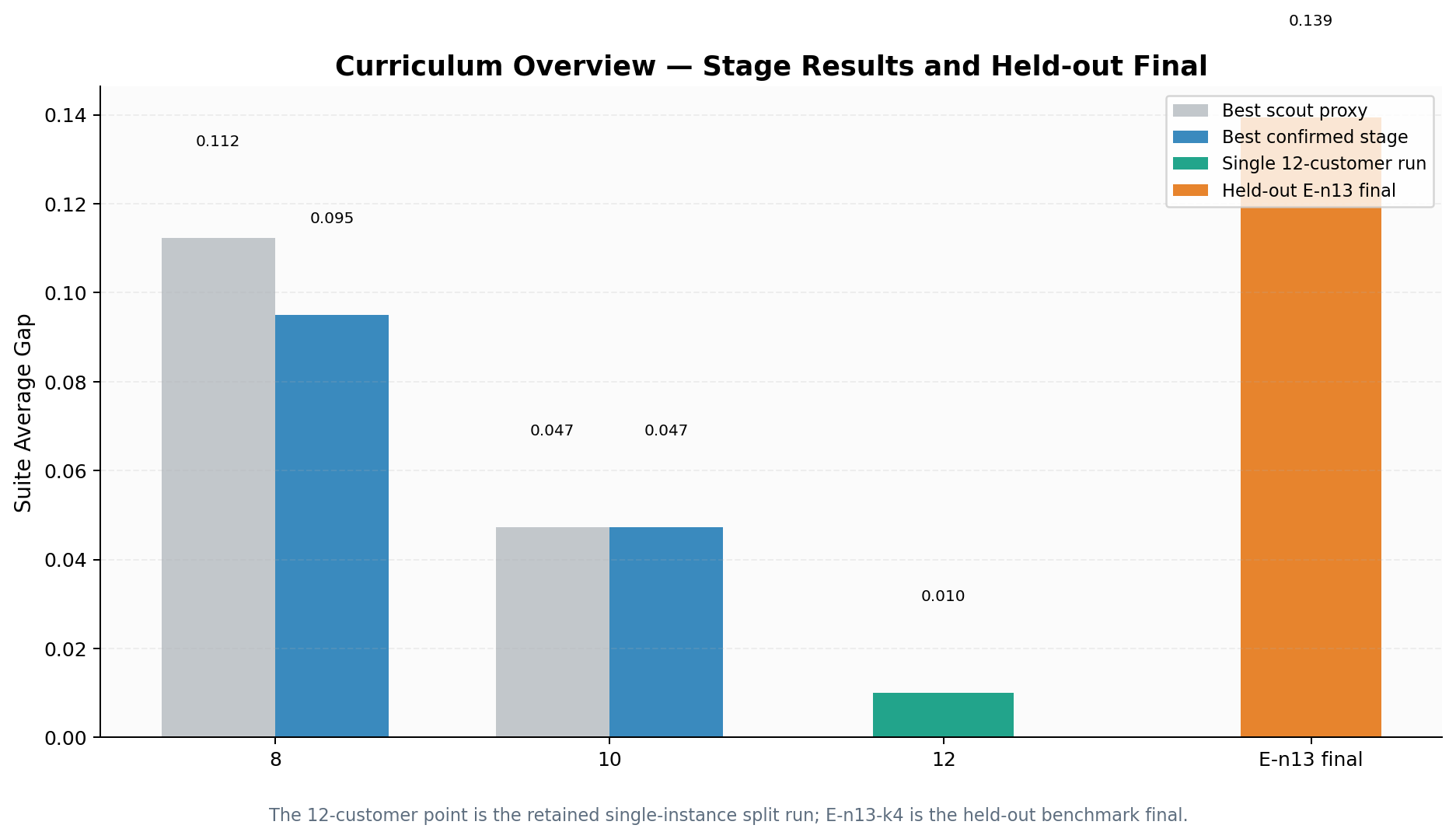}
  \caption{CVRP curriculum overview showing best scout proxy, best confirmed stage,
    retained 12-customer run, and held-out E-n13 final. The strongest gains occur
    between the 8- and 10-customer stages after the move to a more effective penalty
    design, while the held-out benchmark remains more challenging than the curriculum
  stages.}
  \label{fig:cvrp_overview}
\end{figure}

Figure~\ref{fig:cvrp_overview} summarizes the curriculum-level outcomes. The most
important point is that the scout and confirmed stage results align more closely than
in MIS once the tilted-penalty formulation is adopted, suggesting a more stable local
search regime after the main feasibility bottleneck has been resolved. At the same
time, the held-out E-n13 result remains clearly harder than the earlier curriculum
stages, which is consistent with the fact that the routing workflow combines
assignment feasibility and downstream cost quality in a single evaluation target.

\begin{table*}[t]
  \centering
  \caption{Summary of discovered CVRP policies across curriculum stages.}
  \label{tab:cvrp_stage_summary}
  \begin{tabular}{lcc}
    \toprule
    \textbf{Stage} & \textbf{Best confirmed/final gap} & \textbf{Main policy change} \\
    \midrule
    8-customer & 0.095 & higher final sampling \\
    10-customer & 0.047 & tilted penalty + replay-passing sampling \\
    12-customer & 0.010 & retained hybrid single-instance run \\
    E-n13-k4 & 0.139 & held-out final hybrid protocol \\
    \bottomrule
  \end{tabular}
\end{table*}

\subsection{Cross-Problem Lessons}

Taken together, the two case studies support three conclusions. First, the value of
the framework lies in \emph{adaptive policy discovery}, not in selecting a single
globally best solver configuration. Second, the dominant interventions are
problem-dependent: for MIS, the most useful changes are objective selection, solver
family changes, and qubit compression; for CVRP, they are penalty design,
sampling-budget allocation, and workflow-level feasibility control. Third, staged
evaluation is essential. In MIS, proxy evaluations can mis-rank candidates and
materially overestimate performance; in CVRP, replay guardrails are necessary to
prevent gains on a new stage from masking regressions on earlier validated stages.
Together, these results position AutoQResearch as a practical closed-loop
quantum--GenAI workflow for discovering adaptive solver-control policies across
distinct optimization settings.

\section{Executed Hardware Validation}
\label{sec:hardware}

To test whether the discovered policies remain executable beyond simulation, we
performed limited validation runs on IBM quantum hardware for one representative
instance from each problem setting. These runs are not intended as full hardware
benchmarks. Rather, they serve as restricted execution studies showing that the
policies discovered by the closed-loop search can be transpiled, scheduled, and
executed on current devices while preserving useful solution quality.

For MIS, we executed the final 16-node controller on a retained instance using the
same warm-start QAOA policy identified as strongest in simulation. For CVRP, we
executed the retained held-out E-n13-k4 policy using the hybrid controller selected
by the search, in which a reduced quantum assignment subproblem is embedded inside a
larger classical routing workflow. A high-level comparison of the two hardware runs
is reported in Table~\ref{tab:hardware_summary_combined}, while detailed execution
statistics are given in Tables~\ref{tab:mis_hardware_details}
and~\ref{tab:cvrp_hardware_details}.

\subsection{MIS Hardware Validation}

The MIS hardware run used the final stage-1 policy: depth-1 warm-start QAOA with a
CVaR objective (\(\alpha=0.25\)), linear entanglement, COBYLA optimization, and
1024 shots for both estimator and sampler evaluation. The logical circuit used 16
qubits, 2 trainable parameters, logical depth 14, and 22 logical two-qubit gates.
Execution was distributed across \texttt{ibm\_fez} and \texttt{ibm\_marrakesh}, both
with 156 qubits. The total wall-clock time was 369.75\,s. The best sampled
bitstring was feasible and attained the exact optimum value \(8.0\), yielding gap
\(0.0\). At the same time, the measured distribution remained broad:
\texttt{top1\_prob}=0.0322, the raw feasibility rate was \(0.5547\), and 664 unique
bitstrings were observed among 1024 samples. This run therefore shows that the
discovered MIS controller can be executed on current hardware without changing its
solver structure and can still recover an optimal solution on the tested instance,
even though the hardware sampling distribution is far from sharply concentrated.

\subsection{CVRP Hardware Validation}

The CVRP hardware run used the retained E-n13-k4 controller, which is explicitly
hybrid. The policy combines a classical preprocessing step that fixes unambiguous
assignment decisions, a reduced VQE solve for the remaining ambiguous assignment
subproblem, feasibility repair, and classical exact TSP routing on the resulting
clusters. The retained policy uses an \texttt{efficient\_su2} ansatz with one
repetition, linear entanglement, COBYLA, 2048 estimator shots, 16384 sampler shots,
tilted capacity penalties, and depot-farthest seeding. Although the full assignment
model contains 48 variables, the reduced quantum subproblem involves only 16 qubits
because only four customers remain ambiguous after the classical preprocessing step.

The hardware run was executed on the same IBM backends,
\texttt{ibm\_fez} and \texttt{ibm\_marrakesh}, each with 156 qubits. The reduced
quantum circuit used 16 qubits, 64 parameters, total logical gate count 79, and 15
logical two-qubit gates. Representative transpiled circuits on hardware had depth
52--53 with 15 two-qubit gates. The run completed successfully, produced a feasible
repaired assignment, and yielded a routed cost of \(287.0\) against an optimal
reference cost of \(247\), corresponding to gap \(0.139373\). The total
wall-clock time was 4395.93\,s. Importantly, this run improved substantially over
the classical greedy pre-routing assignment inside the same hybrid controller, whose
routed cost was \(311.0\) with gap \(0.205788\). Thus, the executed CVRP controller
was not only hardware-feasible, but also materially useful within the end-to-end
hybrid routing workflow.

Taken together, these two validation studies support the same conclusion from two
different directions. In MIS, the discovered controller transfers to hardware
without losing optimal solution quality on the tested instance. In CVRP, the
discovered controller remains executable even in a richer hybrid workflow that
combines quantum assignment refinement with classical repair and downstream routing.
Although these are restricted single-instance studies and should not be interpreted
as broad hardware benchmarks, they show that the policies discovered by
\textit{AutoQResearch} are not merely simulator artifacts; they can be compiled and
executed on current IBM hardware while preserving nontrivial end-to-end performance.
We stress the methodological limits of these runs: each was executed within a single
calibration window and reported as a single execution, so the numbers reflect one draw
of device conditions (readout noise, routing, and calibration drift) rather than an
averaged or interval estimate. Repeated executions across calibration periods, with
variance or confidence intervals and a study of policy stability under drift, are the
natural next step and are left to future work; the released harness exposes the same
hardware entry points to make such repetition reproducible.

\begin{table*}
  \centering
  \caption{Summary of executed hardware validation runs.}
  \label{tab:hardware_summary_combined}
  \begin{tabular}{lcc}
    \toprule
    \textbf{Field} & \textbf{MIS} & \textbf{CVRP} \\
    \midrule
    Instance & \texttt{1tc.16.txt} & \texttt{E-n13-k4.vrp} \\
    Backends & \texttt{ibm\_fez}, \texttt{ibm\_marrakesh} & \texttt{ibm\_fez}, \texttt{ibm\_marrakesh} \\
    Policy type & WS-QAOA + CVaR & Hybrid VQE + classical routing \\
    Quantum width used & 16 & 16 \\
    Wall time (s) & 369.75 & 4395.93 \\
    Reference objective & 8.0 & 247.0 \\
    Obtained objective & 8.0 & 287.0 \\
    Gap & \(\mathbf{0.0}\) & \(\mathbf{0.139373}\) \\
    Feasible output & Yes & Yes \\
    \bottomrule
  \end{tabular}
\end{table*}

\begin{table*}
  \centering
  \caption{Representative hardware-execution details for the MIS 16-node run.}
  \label{tab:mis_hardware_details}
  \begin{tabular}{ll}
    \toprule
    \textbf{Field} & \textbf{Value} \\
    \midrule
    Policy type & Warm-start QAOA + CVaR \\
    CVaR parameter & \(\alpha = 0.25\) \\
    Entanglement & Linear \\
    Repetitions & 1 \\
    Optimizer & COBYLA \\
    Logical qubits & 16 \\
    Trainable parameters & 2 \\
    Logical depth & 14 \\
    Logical 2Q gates & 22 \\
    Logical total gates & 102 \\
    Estimator shots & 1024 \\
    Sampler shots & 1024 \\
    Backends used & \texttt{ibm\_fez}, \texttt{ibm\_marrakesh} \\
    Total hardware jobs & 24 \\
    Best objective value & 8.0 \\
    Exact optimum & 8.0 \\
    Optimality gap & \(\mathbf{0.0}\) \\
    Feasible best sample & Yes \\
    Raw feasibility rate & 0.5547 \\
    Top-1 probability & 0.0322 \\
    Unique sampled bitstrings & 664 / 1024 \\
    \bottomrule
  \end{tabular}
\end{table*}

\begin{table*}
  \centering
  \caption{Representative hardware-execution details for the CVRP E-n13-k4 run.}
  \label{tab:cvrp_hardware_details}
  \begin{tabular}{ll}
    \toprule
    \textbf{Field} & \textbf{Value} \\
    \midrule
    Policy type & Hybrid reduced VQE + classical exact routing \\
    Assignment penalty & Tilted \\
    Seed method & \texttt{depot\_farthest} \\
    Full assignment variables & 48 \\
    Reduced quantum subproblem qubits & 16 \\
    Ambiguous customers refined quantumly & 4 \\
    Ansatz & \texttt{efficient\_su2}, depth 1 \\
    Optimizer & COBYLA \\
    Estimator shots & 2048 \\
    Sampler shots & 16384 \\
    Logical gate count & 79 \\
    Logical 2Q gate count & 15 \\
    Representative transpiled depth & 52--53 \\
    Representative transpiled 2Q gates & 15 \\
    Classical greedy routed cost & 311.0 \\
    Final routed cost & 287.0 \\
    Optimal reference cost & 247.0 \\
    Final gap & \(\mathbf{0.139373}\) \\
    \bottomrule
  \end{tabular}
\end{table*}

\section{Discussion}
\label{sec:discussion}

\subsection{AutoQResearch as a Closed-Loop Quantum--GenAI Workflow}

The main contribution of this paper is not a new variational quantum algorithm, nor a
new combinatorial formulation, but a closed-loop workflow in which a generative model
proposes adaptive solver-control policies and a quantum optimization stack validates,
rejects, or refines them through measured execution outcomes. This framing is
important because it shifts the role of the LLM from passive assistant to structured
experiment planner. The agent does not merely suggest hyperparameters once; it
observes solver diagnostics, edits a restricted policy surface, and iterates under a
fixed evaluation protocol. In this sense, \textit{AutoQResearch} is best understood
as a quantum--GenAI co-design framework for autonomous solver discovery rather than
as a conventional AutoML tuner.

The two case studies support this interpretation from complementary directions: on MIS
the effective solver family changes with scale (CVaR-enhanced warm-start QAOA at 16
nodes, QRAO-based compression at larger sizes), whereas on CVRP the dominant improvements
are workflow-sensitive (penalty redesign, sampling-budget allocation, hybrid feasibility
repair). That the same closed-loop search remains useful across both regimes is among the
strongest evidence in the paper that the agent navigates distinct optimization regimes
rather than fitting a single problem class.

\subsection{Why Adaptive Policies Matter}

A central lesson is that the appropriate search object is an adaptive policy, not a
single fixed configuration. On MIS the relevant choices are switching measurement
objective, changing solver family, and compression-based fallback; on CVRP they are
changing the assignment penalty, reallocating sampling budget, and embedding the quantum
solver in a hybrid repair-and-routing pipeline. These are qualitatively different
interventions, yet all are expressible within the same restricted interface. This is
where the framework moves beyond static hyperparameter selection: a static configuration
cannot say ``if concentration collapses, move to a compressed solver family'' or ``if
feasibility is the bottleneck, change the penalty and increase final sampling.'' In our
runs this conditional behavior is the mechanism by which the framework improves over the
baseline, supporting the broader claim that sequential policy search is a more natural
formulation of solver design than one-shot configuration search.

\subsection{Methodological Value of Staged Confirmation}

A second major lesson is methodological, and the ablation of Section~\ref{sec:ablation}
makes it concrete: cheap proxy evaluations accelerate search but are not reliable enough
to serve as the sole acceptance rule. The proxy can overestimate stage performance and
even invert candidate rankings---at 32 nodes the confirmed winner was the worst scout-ranked candidate
in its round---while on CVRP replay guardrails are needed to stop apparent gains on a new
stage from masking regressions on earlier ones. This is broader than the specific
benchmarks: any LLM-guided workflow that optimizes against low-cost proxies risks
overfitting them unless it validates on the true target. The scout--promote--confirm
design is therefore part of the paper's methodological contribution, not merely an
implementation convenience.

\subsection{Scope, Limitations, and Future Directions}

The conclusions of this paper should be interpreted within the scope of the studied
framework. First, the action space is curated rather than open-ended, and the
effectiveness of the system derives in part from strong human-designed inductive biases:
the choice of solver families, the diagnostic signals, and the promotion protocol are all
engineered. The framework is therefore more accurately described as \emph{structured
adaptive policy search} than as open-ended autonomous scientific discovery, and the
discovered policies are optimal only relative to the explored design space. We use
``discovery'' throughout in this bounded sense---discovery of adaptive policies within a
curated space, not invention of new algorithms.

Second, and most importantly, we do not yet isolate the contribution of the LLM from that
of the curated surface and staged harness. The paper provides a log-derived
staged-confirmation ablation (Section~\ref{sec:ablation}), search-cost and
LLM-reliability accounting (Section~\ref{sec:agent_config}), and a full specification of
the agent (model, prompt, protocol), but these analyses substantiate the workflow rather
than separate the agent's contribution from the interface. What remains open is a \emph{matched-budget head-to-head} against random search,
Bayesian optimization, evolutionary search, and bandit controllers \emph{over the
identical policy surface}. Because the policy space is a discrete, conditional program
space rather than a smooth parameter vector, such a comparison requires care in adapting
each baseline to the surface; the released harness exposes the frozen-baseline and
study-runner entry points precisely so this comparison is reproducible, and we consider it
the single most valuable follow-up.

Third, the empirical scale and statistical robustness remain limited: the MIS 48- and
64-node results rest on retained sparse instances, the CVRP 12-customer and held-out
hardware studies are single-instance evaluations, and search is driven from a fixed seed.
We report per-instance gaps and provide a post-lock five-seed robustness protocol, but we
do not report variance or confidence intervals across many independent agent trajectories;
larger instance suites, multiple independent runs, and interval estimates would make the
empirical claims more conclusive. Finally, policies are discovered under idealized
noiseless simulation, and the hardware validation---while confirming executability---should
not be read as a broad, noise-averaged hardware benchmark.

These limitations suggest several natural directions for future work. One is to
expand the benchmark set so that the closed-loop search can be evaluated across a
larger family of optimization workflows and hardware backends. A second is to compare
LLM-guided policy search directly against other structured search methods under a
common interface and budget. A third is to expose richer policy surfaces, including
resource-aware routing across heterogeneous backends or learned policies for when to
invoke quantum versus classical subroutines inside hybrid workflows. More broadly,
the results here suggest that the most promising role for generative AI in quantum
optimization may not be unrestricted algorithm invention, but disciplined closed-loop
search over structured solver-design spaces.

\section{Open-Source Release}
\label{sec:open_source}

To support reproducibility, we release \textit{AutoQResearch} as a public artifact
accompanying this paper. The repository includes the core framework, solver
implementations for VQE, QAOA, QRAO, and PCE, benchmark instances used in the study,
and the policy-search infrastructure required to reproduce the reported experiments.

The release also provides experiment artifacts generated during the search process,
including promoted policy checkpoints, execution logs, the instruction file
(\texttt{program.md}), hardware-execution specifications, and analysis scripts for
regenerating the main plots and tables. To support exact reproduction, the release documents the driving model
(GPT-5.3-Codex) and the full unmodified prompt, the fixed evaluation seed and the
post-lock seed list, the simulation backend configuration (\texttt{ideal\_mps} /
statevector with the shot budgets used), the per-stage instance membership and splits,
and the complete policy-edit history---every proposed \texttt{experiment.py} diff with its
keep/revert decision---for both accepted and rejected candidates. These materials are
intended to make the closed-loop search process auditable rather than only its final
outputs.

Beyond reproducing the reported results, the released artifact supports follow-up
work on autonomous quantum optimization workflows. In particular, it enables
controlled re-evaluation of discovered policies, analysis of how prompt and protocol
design affect search behavior, and extension of the framework to additional problem
families and backend settings.

\section{Conclusion}

We presented \textit{AutoQResearch}, an LLM-guided closed-loop framework for adaptive
variational quantum optimization. Rather than treating solver design as static
hyperparameter tuning, the framework casts it as sequential policy search over a
curated and auditable solver space, in which a generative model proposes adaptive
solver-control logic and a fixed quantum evaluation harness validates, rejects, or
refines those proposals through measured outcomes.

Across two structurally different optimization settings, Maximum Independent Set
(MIS) and a decomposed Capacitated Vehicle Routing Problem (CVRP) workflow, the
framework discovered nontrivial problem- and scale-dependent policies that improved
substantially over their respective baselines: a transition from CVaR-enhanced
warm-start QAOA to QRAO-based compression on MIS, and penalty-, sampling-, and
repair-level adaptations on CVRP\@. The value of the framework lies not in selecting one
universally best configuration, but in discovering policies that adapt to the dominant
failure mode of the current regime, and the consistency of gains across both problems
suggests structure-aware adaptation rather than problem-specific heuristics.

A second key finding is methodological: staged confirmation is essential for credible
autonomous experimentation. As the ablation of Section~\ref{sec:ablation} shows on real
promotion logs, cheap scout evaluations can misestimate policy quality and even invert
candidate rankings, so the scout--promote--confirm protocol with replay guardrails is
part of the contribution, not an implementation detail.

Overall, this work positions \textit{AutoQResearch} as a benchmarked
quantum--GenAI co-design workflow for autonomous solver discovery. More broadly, the
results suggest that a promising role for generative AI in quantum optimization is
not unrestricted algorithm invention, but disciplined closed-loop search over
structured solver-design spaces, where candidate policies are proposed by AI and
validated by quantum execution. Extending this framework to additional benchmark
families (e.g., portfolio optimization and network design) is a natural next step.

\section*{Acknowledgements}
This project is supported by the National Research Foundation, Singapore through the National Quantum Office, hosted in A*STAR, under its Quantum Engineering Programme 3.0 Funding Initiative (W24Q3D0002).

\balance
\bibliography{reference}

\end{document}